1/12/04

# Multimer Radical Ions and Electron/Hole Localization in Polyatomic Molecular Liquids: A critical review.

Ilya A. Shkrob and Myran C. Sauer, Jr.

*Chemistry Division , Argonne National Laboratory,  Argonne, IL 60439*

**Abstract.**

While ionization of some polyatomic molecular liquids (such as water and aliphatic alcohols) yields so-called "solvated electrons" in which the excess electron density is localized in the interstices between the solvent molecules, most organic and inorganic liquids yield radical anions and cations in which the electron and spin densities reside on the solvent molecule or, more commonly, a group of such molecules. The resulting multimer ions have many unusual properties, such as high rates of diffusive hopping. The "solvated electron" can be regarded as a variant of a multimer radical anion in which the charge, while perturbing the solvent molecules, mainly resides in the space between these molecules. We give several examples of less known modes for electron localization in liquids that yield multimer radical anions (such as $C_6F_6$, benzene, acetonitrile, carbon disulfide and dioxide, etc.) and holes localization in liquids that yield multimer radical cations (such as cycloalkanes). Current understanding of the reaction properties for these high-mobility solvent radical anions and cations is discussed.

**I. Introduction**

Interaction of ionizing radiation – fast electrons, α-particles, x- and γ- rays, and UV and VUV photons - with molecular solids and liquids causes the formation of short-lived electron-hole pairs that, in such media, thermalize and, eventually, localize yielding radical ions and/or trapped (solvated) electrons and holes. The distinction between the radical anions and the solvated electrons is arbitrary. For the time being, it will be assumed that "radical ions" have an excess electron or electron deficiency in the valence orbitals of a single solvent molecule ("molecular ions", "monomer ions") or a small group of such molecules ("dimer ions" or "multimer ions") that do not share charge with neutral solvent molecules that "solvate" them. Naturally, the excess electron in a "radical anion" is indistinguishable from other valence electrons in this anion. By contrast, in the "solvated electron" (also known as " cavity electron"), the electron density resides mainly in interstitial sites between the solvent molecules ("solvation cavity") that are polarized by the negative charge at its center (thereby forming the outer shell of a "negative polaron"). The underlying assumption of this visualization is that the "solvated electron" is a single-electron state whose properties can be given by a band model in which the valence electrons in the solvent and the excess electron in the cavity are treated wholly separately [1] - in the exact opposite way to how the electronic structure of the solvent "radical ion" is viewed. An additional assumption is usually made that the excess electron interacts with (rigid, flexible, or polarizable) solvent molecules by means of an empirical *classical* potential. Both of these simplifying assumptions find little support in structural studies of "trapped electrons" in vitreous molecular solids using magnetic resonance spectroscopy [2].

The "primary" species – solvated/trapped electrons/holes and solvent "radical ions" - are efficient donors and acceptors of electrons and protons; they readily react with the solvent, the solute and dopant molecules, each other, and the short-lived species (radicals, molecular fragments and excited states) generated in the ionization and excitation events along with these "primary" charges. In most radiation chemistry studies, the species of interest are the resulting "secondary" ions, radicals, and excited neutrals.

Quite often in such studies, the radiolysis is complemented by other techniques for radical ion generation, such as plasma oxidation, electron bombardment – matrix deposition, and chemical and electrochemical reduction or oxidation.

There are many excellent books and reviews on the structure and reactions of *secondary* radical ions generated in radiolytic and photolytic reactions. Common topics include the means and kinetics of radical ion production, techniques for matrix stabilization, electronic and atomic structure, ion-molecule reactions, structural rearrangements, etc. On the other hand, the studies of *primary* radical ions, viz. solvent radical ions, have not been reviewed in a systematic fashion. In this paper, we attempt to close this gap. To this end, we will concentrate on a few better characterized systems (there have been many scattered pulse radiolysis studies of organic solvents; most of these studies are inconclusive as to the nature of the "primary" species).

Before we review specific systems, note that the *primary* species should be considered on a different footing than the *secondary* radical ions. The latter ions are well-isolated from their parent molecules by the matrix or the solvent. By contrast, in the *primary* species, the charge is residing on a molecule(s) that is surrounded by like solvent molecules. This often results in unusual properties because the barriers for charge hopping and charge delocalization are lowered. We will examine several examples of the reaction and migration dynamics of such *primary* radical ions; the multiplicity of examples and commonality of the observed behavior suggest a general pattern.

As demonstrated below, a primary charge viewed as a solvated electron or the molecular ion residing in an inert liquid does not account for experimental observations in many if not most of the systems. While we cannot offer a specific, general model of these "exceptional" ions, we provide a general introduction to the known properties of such species. Furthermore, we argue that these species comprise the rule rather than the exception. The reader is invited to reach his or her own conclusions.

## II. Electrons and solvent anions in supercritical $CO_2$.

The first "exceptional" system that we review is carbon dioxide [15-21]. Supercritical (sc) $CO_2$ finds numerous industrial applications as a "green" solvent, and this practical consideration stimulates interest in its radiation chemistry. Though the studies of sc $CO_2$ are recent, this system is particularly interesting because of the simplicity of the solvent molecule and extensive gas phase and matrix isolation studies of the corresponding ions (see below).

Like other sc fluids, sc $CO_2$ is a collection of nanoscale molecular clusters that rapidly associate, dissociate, and exchange molecules among each other [22]. Pre-thermalized electrons in sc $CO_2$ attach to these $(CO_2)_n$ clusters. In the gas phase, the attachment can be dissociative (reaction (2)) and or non-dissociative (reaction (1)),

$$e^- + (CO_2)_n \rightarrow (CO_2)_n^- \qquad (1)$$

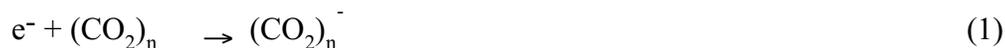

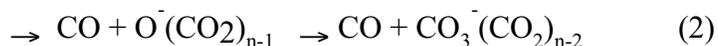

$$\rightarrow CO + O^-(CO_2)_{n-1} \rightarrow CO + CO_3^-(CO_2)_{n-2} \qquad (2)$$

depending on the electron energy (for $n=2$, reaction (2) is 3.6 eV more endothermic) [23]. Although in radiolysis, a large fraction of electrons have initial excess energies > 10 eV, the yield of $CO_3^-$ in 20 MeV electron radiolysis of dense sc-$CO_2$ does not exceed 5% of the total ionization yield [20].

In the gas phase, anions formed in reaction (1) have been extensively studied [24]. A linear $CO_2$ molecule has negative gas-phase electron affinity ($EA_g$) of -0.6±0.2 eV. The metastable $C_{2v}$ monomer anion, $CO_2^-$ (with OCO angle of 135° and autodetachment time of < 100 µs) exhibits vertical detachment energy (VDE) of 1.33 to 1.4 eV. This energy increases to 2.6-2.79 eV in the stable, $D_{2d}$ symmetric C-C bound dimer anion, $C_2O_4^-$, shown in Fig. 1a (with lifetime > 2 ms), and further increases to 3.4 eV for $n=6$ clusters. In larger clusters, the VDE first decreases to 2.49 eV ($n=7$), then monotonically increases to 3.14 eV ($n=13$), then, for $n \geq 14$, the VDE suddenly increases to 4.55 eV. The onset of the photoelectron spectrum increases from 1.5 eV for $n=2-7$ clusters to 1.8-2 eV for $n=8-13$ clusters to 3 eV for $n=15-16$ clusters. Tsukuda et al. [24] argue that while the core of small ($n \leq 6$) and large ($n \geq 14$) clusters is a $D_{2d}$ dimer anion, the core of $6<n<14$ clusters is a monomer anion weakly coupled to several $CO_2$ molecules (with binding energy ca. 0.22 eV per molecule). Both of these forms coexist in $n=6$ and $n=13$ clusters. It appears that the two isomers are close in energy, and core switching occurs readily when the cluster geometry and size change. Since in sc fluid there is a wide distribution of the cluster sizes and shapes [22], the solvent radical anion does not have a well-defined geometry.

In $CO_2$ gas, the density-normalized electron mobility $\mu_e \tau_e$ is independent of temperature ($2 \times 10^{22}$ molecule/cm·V·s [25]), though the apparent mobility steadily decreases with the pressure: free electrons are trapped by neutral $(CO_2)_n$ clusters ($n \approx 6$) with nearly collisional rates, and the electron affinity of these clusters > 0.9 eV. When extrapolated to solvent densities of $(2-15) \times 10^{21}$ cm$^{-3}$ typical for sc $CO_2$, these estimates suggest that the free electron mobility $\mu_e$ is ca. 1 cm$^2$/Vs and its collision-limited lifetime $\tau_e$ < 30 fs [18]. If the lifetime were this short, the electrons would contribute negligibly either to the conductivity or the product formation. However, this extrapolation is not supported by experiment [18,20].

In low-temperature solid matrices (e.g., Ne and Ar at 5-11 K), the $CO_2^-$ monomer, $(CO_2^-)(CO_2)_n$ multimer anion ($n = 1, 2$), and the $C_2O_4^-$ dimer were observed using IR spectroscopy; the latter species was the only one stable at 25-31 K [26]. In nonpolar liquids, monomer $CO_2$ is an efficient electron scavenger. Still, the electron can be detached from the monomer anion both thermally and photolytically [27]. For instance, in *iso*-octane, reaction (1) is exothermic by 1.08 eV, the VDE peak is at 3 eV, and the photodetachment threshold is at 1 eV [27].

As seen from the above, the mode of electron trapping in sc-$CO_2$ cannot be deduced from the results obtained in the gas phase or matrix isolation studies. It is not obvious whether the solvent radical anion should be similar to multimer cluster anions found in the gas phase, dimer cation(s) in solid matrices, or monomer $CO_2^-$ anions in inert liquids. Such a situation is typical for other molecular liquids.

Time-resolved laser dc photoconductivity and pulse radiolysis - transient (electro)absorbance studies of sc-$CO_2$ showed that ionization of the solvent (or UV-light absorbing solute) yields two negatively charged species: a metastable (quasifree) conduction band (CB) electron and a rapidly hopping multimer anion (a self-trapped electron) [18,20]. Both of these species exhibit unusual properties that account for many oddities observed in radiolysis of sc $CO_2$. Quasifree electrons in sc $CO_2$ have lifetime $\tau_e <$ 200 ps and mobility $\mu_e > 100$ cm$^2$/Vs [18]. For reduced solvent density $\rho_r > 1.2$ (defined as the ratio of the solvent density $\rho$ and the critical density $\rho_c = 0.468$ g/cm$^3$) the product $\rho_r\mu_e\tau_e$ exponentially increases with $\rho_r$. The onset for the formation of rapidly migrating quasifree electrons coincides with the emergence of the CB in the solvent [18]. Exactly the same behavior was observed by Holroyd and co-workers for sc saturated hydrocarbons. Their studies suggest that the electron mobility increases exponentially with density between $\rho_r = 1$ and 2 and then stabilizes and/or slightly decreases at greater density [28]. In sc $CO_2$, the product $\mu_e\tau_e$ shows signs of saturation at the reduced density of 1.8 (reaching ca. $2.5 \times 10^{-9}$ cm$^2$/V) [18]. Both this behavior and the high mobility indicate that the quasifree electron is not attached by $CO_2$ clusters, even temporarily, before it is finally trapped. This is surprising, given the extremely rapid rate of electron attachment to $(CO_2)_n$ clusters in the gas phase. Apparently, once the CB is formed in a dense liquid, the electron dynamics changes dramatically.

In sc $CO_2$, both the solvent viscosity and the mobility of solute ions (e.g., halide anions and aromatic and alkylamine cations) are a function of the solvent density rather than the solvent temperature and pressure separately [18,21]. In other words, if the density is constant, the ion mobilities do not change with the solvent temperature. At a given temperature, the ion mobility decreases rapidly with $\rho_r$ for $0.2 < \rho_r < 1$ and then decreases slowly for $1 < \rho_r < 2$ [18,21]. In contrast to this behavior, the mobility of the solvent anion exponentially increases with $\rho_r$, being 2-10 times greater than the mobilities of all other ions in sc $CO_2$ [18]. The activation energy of the solvent anion migration is 0.46 eV (for constant $\rho_r$) whereas for the solute ions (Fig. 1c), this energy is less than 20 meV [18,21]. Careful analysis of the data on the solvent anion and electron dynamics and thermochemistry indicate that an equilibrium between the quasifree and trapped electrons, similar to that observed in saturated hydrocarbons, cannot account for the observed dynamics. This is reasonable, given that the trapping energy of the electron is almost an order of magnitude larger in sc $CO_2$ than in these hydrocarbons.

Anomalously high mobility and large activation energy for migration of the solvent anion suggests that this anion migrates by rapid charge hopping between the solvent clusters (Fig. 1b); this hopping easily outruns Brownian diffusion of the core anion [18]. The hopping mechanism is also suggested by the fact that the mobility exponentially increases with $\rho_r$, at any temperature: as the solvent density increases, the cluster-to-cluster distance decreases, and the coupling integral becomes greater; the hopping rate increases accordingly.

Electron photodetachment upon laser excitation of the solvent anion above 1.76 eV was observed (Fig. 2a,c) [18]. The cross section of photodetachment linearly increases between 1.76 and 3 eV (Fig. 2b). Under the same physical conditions, the photodetachment and absorption spectra of the solvent anion are identical (Fig. 2b) [20], suggesting a bound-to-CB transition; the quantum yield of the photodetachment is close to unity. The photodetachment spectrum is similar to the photoelectron spectra of $(CO_2)_{6-9}^-$ clusters observed by Tsukuda et al. in the gas phase [24]; it is distinctly different from the electron photodetachment spectra of $CO_2^-$ in hydrocarbon liquids [27]. This suggests that a C-C bound, $D_{2d}$ symmetric dimer anion constitutes the core of the solvent radical anion [18,19].

Both the electrons and the solvent anions react with nonpolar electron acceptor solutes with $EA_g > 0.4$ eV [18]. For oxygen ($EA_g$ of 0.4 eV), the electron transfer from the solvent anion to the solute (which yields a stable $CO_4^-$ anion) is reversible (Fig. 3); the free energy of the corresponding reaction is -0.42 eV. The rates of the electron attachment and solvent anion scavenging correlate with each other and the $EA_g$. On the other hand, the correlation of these rate constants with the free energy ($\Delta G^o$) of the overall reaction (expected from Marcus' theory of electron transfer) is very poor [18]. The same pattern is observed for high mobility solvent *cations* in nonpolar liquids (section V). The reason for this behavior is that while the overall $\Delta G^o$ depends on what happens to the products (e.g., solute ions) *after* the electron transfer (structural relaxation, solvation, fragmentation, and bonding to the solvent), the activation energy of the reaction depends only on the vertical electron affinity (ionization potential) of the solvent anion (cation) and the solute [18]. This can be rationalized assuming that the scavenging reaction occurs by *direct* electron transfer to (or from) the solute: if the solute or solute needs "reorganization" in order to accept (donate) the electron, the reaction does not occur and the solvent ion migrates away from the solute molecule. In other words, due to the extremely fast charge hopping, the lifetime of the collision complex is always shorter than the time needed for stabilization of the solute ion by solvation and/or structural relaxation. No such anomalies are observed for (relatively slow) electron transfer reactions that involve *regular ions* in the same nonpolar liquids, and those do conform to the Marcus theory (though, to our knowledge, the inverted region has never been observed in such reactions). For example, rate constants for charge-transfer reactions of solvent *holes* in sc $CO_2$ with electron donors (CO, $O_2$, $N_2O$, and dimethylaniline) systematically increase with the ionization potential of the solute [15,16]. These solvent holes are $C_{2h}$ symmetrical O-O bound $(CO_2)_2^+$ cations that exhibit a prominent charge resonance band in the visible, which suggests strong coupling in the dimer [15,16,19,20]. These solvent cations exhibit normal diffusion properties and rate constants of diffusion-controlled electron transfer reactions [15,16,20]; no ultrafast charge hopping is suggested by the data [20]. Such a situation is common: no known liquid or solid yields both high-mobility anions *and* high-mobility cations. The likely reason is that at least one of these species has a tendency to form strongly bound dimer radical ions that cannot migrate rapidly by the hopping mechanism (see below).

In sc $CO_2$, only solutes with $EA_g > 2$ eV exhibit diffusion-controlled kinetics for reaction with the solvent anion (which is consistent with the electron trapping energy between 1.6 and 1.8 eV that was estimated from the photodetachment spectrum) [18]. For $\rho_r > 0.85$, the scavenging radii of these diffusion-controlled reactions systematically decrease with the solvent density, $\rho_r$ [18]. Interestingly, in addition to the electron transfer reactions, solvent anions in sc $CO_2$ form 1:1 and 1:2 complexes with polar molecules that have large dipole moments, such as water, aliphatic alcohols, alkyl halides, and alkyl nitriles [29]. None of these polar low-$EA_g$ solutes directly reacts with quasifree electrons in sc $CO_2$. The complexation rate is 10-50% of the diffusion controlled rate, the equilibrium constants of the 1:1 complexation range from 10 to 350 $M^{-1}$ depending on the solute, the reaction heat is - 15 to -21 kJ/mol, and the reaction entropy is negative [29]. The stability of these complexes increases with the dipole moment of the polar group and decreases with substitution at the α-carbon. It appears that these complexes are bound by weak electrostatic interaction of the negative charge and the molecular dipole. Previously, such dipole-bound complexes, with monomers of acetonitrile [30] and dimers and higher multimers of aliphatic alcohols [31], were observed for "solvated electrons" in saturated hydrocarbons. For these electrons, the complexation manifests itself through a precipitous decrease in the electron mobility upon the addition of the solute: thermal emission of trapped electrons from the complexes to the CB is much less efficient than that of free "solvated electrons". No changes in the absorption band of the "solvated electron" in the NIR are observed upon the complexation of these electrons with the monomers and dimers of polar molecules [32]. Higher alcohol multimers, such as tetramers, provide very deep traps for these electrons; absorption spectra of such cluster-trapped electrons are almost identical to the spectra of "solvated electrons" in neat aliphatic alcohols [33].

That the solvent anion in sc $CO_2$ demonstrates behavior similar to that of "solvated electrons" in saturated hydrocarbons suggests that the mechanism of binding to the polar molecules must be similar. Apparently, increasing the trapping energy by ca. 0.15 eV due to the electrostatic binding to a monomer or dimer molecule with a dipole moment of (2-4) Debye is sufficient to halt the charge hopping completely [29]. The same effect of polar solutes on the hopping rate of high-mobility solvent radical *cations* (with fairly similar thermochemistry) was observed in neat *cis*- and *trans*-decalin (section V).

Sc $CO_2$ is not the only liquid for which high-mobility solvent anions are observed (see below), but it is the simplest one. In monoatomic liquids (liquefied Ar, Kr, and Xe), ammonia, and simple hydrocarbons ($CH_4$, $C_2H_6$), solvent anions are not formed (quasifree or "solvated electrons" are formed instead). In diatomic liquids, such as $N_2$ and $O_2$, solvent anions are generated, but the formation of strongly-bound dimer anions precludes rapid electron hopping. Even for $CO_2$, high mobility anions are not observed in the low-temperature liquid: considerable thermal activation is needed to break the dimer anion prior to every hop of the electron. In liquids whose solute molecules show less tendency to form strongly bound dimer anions, the electron hopping is faster and requires

1/12/04

much less activation energy. As argued below, such a situation occurs in carbon disulfide, hexafluorobenzene, benzene, and toluene.

### III. Solvent anions in liquid $CS_2$, $C_6F_6$, and aromatic hydrocarbons.

Carbon disulfide is isovalent to carbon dioxide and it also has a bent monomer anion. While gas-phase $CO_2$ has negative $EA_g$ of –0.6 eV [24], for $CS_2$, $EA_g$ is +0.8 eV [34]. Despite this very different electron affinity, Gee and Freeman observed long-lived "electrons" in $CS_2$ (with life time > 500 μs) with mobility ca. 8 times greater than that of solvent cations [34]. Over time, these "electrons" converted to secondary anions whose mobility was within 30% of the cation mobility. Between 163 and 500 K, the two ion mobilities scaled linearly with the solvent viscosity, as would be expected for regular ions. Of course, Gee and Freeman's identification of the long-lived high mobility solvent anions as "electron" is just a manner of speech: obviously, quasifree or "solvated" electrons cannot survive for over a millisecond in a positive-$EA_g$ liquid.

To the best of our knowledge, pulse radiolysis – transient absorption studies of neat $CS_2$ have not been reported. "$CS_2^-$ anion" in 0.1 M cyclohexane and 0.1 M THF solutions appears as a single 275 nm peak [35]; there is no charge-resonance band that can be attributed to the dimer anion, at early (< 10 ns) or later times.

The studies carried out in the gas phase and low-temperature matrices suggest that $(CS_2)_n^-$ anions have somewhat different structure from the $(CO_2)_n^-$ anions discussed in the previous section [24]. Similarly to the $(CO_2)_n^-$ anions, the $CS_2^-$ monomer and C-C bound $C_2S_4^-$ dimer anions are switching as the core of the $(CS_2)_n^-$ anion [36]. However, unlike the two core anions in the $(CO_2)_n^-$ anions, $CS_2^-$ and $C_2S_4^-$ anions coexist in the clusters of all sizes (*n*=2-6) with the monomer core being statistically prevalent [36]. The dimer core is responsible for the 1.5-1.8 eV peak observed in the photodestruction spectra of *n*=2-4 anions; no matching peak is observed in the liquid [37]. The dimer anion has either $D_{2h}$ (C-C bound) or $C_{2v}$ symmetry (C-C and S-S bound) rather than the $D_{2d}$ symmetry of the C-C bound $C_2O_4^-$ anion [36,37,38]. It is almost certain that the S-S bound structure is energetically preferable, because the UV-vis photoexcitation of $C_2S_4^-(CS_2)_{n-2}$ anions results in the fragmentation of the dimer core to $C_2S_2^-$ and $S_2$ [37]. The dimer anion was also observed in the EPR spectra of γ-irradiated $CS_2$-doped frozen alcohols; the *g*-tensor parameters suggested S-S bonding [39]. Interestingly, while $C_2O_4^-$ is readily observed using IR spectroscopy when electrons are injected in frozen $CO_2$/Ne or $CO_2$/Ar mixtures at 4 K [26], only $CS_2^-$ is observed in similar mixtures doped with $CS_2$ [40]. It seems that dimerization of the molecular anion in $CS_2$ is less favored than in $CO_2$, both in the gas and solid phase - perhaps, due to the positive electron affinity of the monomer molecule.

Despite these structural differences, both liquid $CS_2$ and sc $CO_2$ yield high-mobility solvent anions whose mobilities are similar. The most striking difference is the activation energy for the anion migration. For $CS_2$, this activation energy is only 5 kJ/mol [34] whereas for $CO_2$ it is 46 kJ/mol [18]. Such a large difference is surprising because similar transport mechanisms were suggested for both of these anions. This result becomes more understandable when other examples of high mobility anions are examined.

A relatively long-lived (> 100 ns) high-mobility solvent anion has been observed by microwave conductivity in room-temperature β-irradiated liquid hexafluorobenzene ($C_6F_6$) [41]. Like $CS_2$, hexafluorobenzene has positive $EA_g$ estimated to be between 1 and 2 eV [41]. The anion mobility is 40 times greater than that of the solute ions and the activation energy for the solvent anion migration is 0.11 eV [41]. The electron in the solvent anion is strongly bound: the anion does not react with such efficient electron acceptors as $SF_6$, although it reacts with $CBr_4$, $CCl_4$, and $(NC)_2C=C(CN)_2$ (with rate constant as large as $1.5 \times 10^{11}$ $M^{-1}$ $s^{-1}$). Addition of small amounts of inert solvents (benzene, saturated hydrocarbons, $C_6F_{12}$) results in the exponential decrease in the anion mobility with the molar fraction x of the inert solvent [41,42]. E.g., addition of 5 mol % of cyclohexane drops the mobility by 50% relative to neat $C_6F_6$ [41]. This decrease can be approximated by $(1-x)^n$, where the exponent $n$=15-20. A "percolation" model of charge migration [42] suggests that the negative charge in $C_6F_6$ is spread over ca. 12 solvent molecules; this is why even slight dilution has strong effect. The multimerization is also consistent with the emergence of a 675 nm anion band (with molar extinction coefficient of 5,000 $M^{-1}$ $cm^{-1}$) in pulse radiolysis of neat $C_6F_6$ [41]. This band is different from the $C_6F_6^-$ band observed in dilute solutions of $C_6F_6$ in inert liquids; the latter is centered at 480 nm [41]. Thermochemistry considerations suggest that the 675 nm band cannot be from a sandwich dimer anion similar to the dimer cations of benzene and other planar aromatic hydrocarbons [43]. This conclusion is consistent with optically-detected magnetic resonance (ODMR) and magnetic level-crossing data that indicate that in cold hydrocarbon solutions, the encounter of $C_6F_6^-$ with $C_6F_6$ results in diffusion-controlled degenerate electron transfer rather than anion dimerization [44]. Thus, while the data clearly point to charge sharing and charge hopping, there seems to be no evidence that a metastable dimer is formed. Once more, one needs to postulate a flexible-structure multimer anion in which fractional negative charge freely exchanges between the solvent molecules.

Importantly, both for liquid $C_6F_6$ and $CS_2$, there is a relatively high yield of *ortho* positronium (*o*-Ps) observed in $e^+$ irradiation of the fluids [34]. The *o*-Ps is formed in the $e^+e^-$ recombinations that occur in the end-of-track spurs. The higher is the negative charge mobility, the higher is the probability that these $e^+e^-$ recombinations occur before the $e^+$ undergoes pick-off annihilation with an electron bound in a solvent molecule. Following Gee and Freeman [34], we suggest that high *o*-Ps yield is a general indicator for the presence of high-mobility solvent anions in molecular fluids with positive $EA_g$.

Yet another example of a high-mobility anion is given by benzene and toluene [45], whose molecules have *negative* $EA_g$ of -1 eV. In dilute solutions of benzene and toluene in saturated hydrocarbons and tetramethylsilane, there is an equilibrium between the "solvated electrons" and benzene/toluene anions [45]. This equilibrium is shifted towards the anion *($\Delta G^o$ of - 9 kJ/mol)*, and this shift becomes greater at higher pressure. At low pressure, the data for neat liquid benzene and toluene can be interpreted the same way. However, at high pressure (1-2 kbar), the negative charge mobility becomes independent of pressure, which indicates that no volume change occurs during the charge migration [45]. Any mechanism that requires thermal emission of the electron from the bound state back into the CB would require such a change; only resonant charge transfer can account for the zero reaction volume. That the electron in pressurized benzene is not quasifree also follows from the extrapolation of medium-pressure equilibrium constants to the high-pressure range. These estimates give an equilibrium constant of 10-100 for the bound electron. Simple calculation shows that the estimated fraction of free electrons is too small to account for the observed negative charge mobility. Interestingly, the activation energy for anion hopping in liquid benzene and toluene (0.12 and 0.13 eV, respectively [45]) is very close to that in hexafluorobenzene (0.11 eV [41]), despite a large difference in their $EA_g$; the corresponding anion mobilities are also comparable. Liquid benzene and toluene provide the only known examples of a pressure-induced switch from the thermally-activated electron detrapping to charge hopping.

As in the case of hexafluorobenzene solvent anion, EPR and ODMR spectroscopies suggests that no dimerization of monomer radical anions of benzene and toluene occur in liquid benzene and/or in alkane solutions of benzene (whereas the radical *cation of* benzene is known to dimerize rapidly). The conductivity studies also indicate that there is no volume change associated with the dimerization [45].

We conclude this section with the following observations:

First, high mobility anions occur both in liquids whose molecules have negative and positive $EA_g$. The gas phase electron affinity has no effect on the rate and the activation barrier of electron hopping in neat liquid solvents.

Second, the formation of strongly bound dimer anions is detrimental for rapid charge hopping. Indeed, the dimer must dissociate every time the negative charge moves; this requires thermal activation. As discussed in section II, high-mobility anions in sc $CO_2$ have dimer radical anions as their chromophore core; this only results in a higher activation barrier for hopping and a moderate increase in the anion mobility relative to that of solute ions (a factor of two at the critical temperature). By contrast, high-mobility solvent anions in liquid $CS_2$, hexafluorobenzene, benzene, and toluene (for which the tendency for anion dimerization is weak), have 3-5 times lower activation barriers for the charge hopping and substantially higher migration rates (up to 15 times) than the solute ions. We speculate that the only reason why high mobility solvent anions are observed in sc $CO_2$ at all is the fact that the core rapidly switches between the monomer and dimer anion with change in the cluster size.

Third, charge delocalization over many solvent molecules, perhaps as many as 10-15, seems to be the only way to explain the observations (such as the effect of the dilution on the conductivity and the emergence of new absorption bands in the UV-vis spectra). Classifying these solvent anions as molecular ions "solvated" by their parent liquids or "solvated electrons" does not explain these properties.

It may appear from the above that only *nonpolar* liquids yield "non-molecular" solvent anions upon the ionization. Perhaps this is misleading: Most polar liquids studied by radiation chemists are aliphatic alcohols and water, and these liquids yield "solvated electrons" rather than radical ions. Though there has been sporadic interest in other polar liquids (e.g., neat acetone), the current state of knowledge of such systems does not allow one to reach *any* conclusion as to the nature of the reducing species observed therein (although, see section IV).

It may also appear that the few nonpolar liquids considered above do not comprise the rule. Again, we stress that the most extensively studied *nonpolar* polyatomic liquids are saturated hydrocarbons; it so happens that these fluids also yield "solvated electrons" (that are in a dynamic equilibrium with quasifree CB electrons). Actually, very few nonpolar liquids other than alkanes and liquefied rare gases have been studied by pulse radiolysis or photoconductivity. In almost all such systems, either the "hole" or the "electron" exhibit unusual migration or reaction dynamics that are suggestive of rapid charge hopping. Given that in many liquids the primary solvent ions are short-lived, we hazard a conjecture that most liquids which do not yield "solvated electrons" yield high-mobility, multimer solvent anions or (as shown in section V) solvent holes. The true scope for the occurrence of such ions is not known, but the number of examples steadily increases.

## IV. Solvent anions and electron localization in liquid acetonitrile.

The previous examples of high mobility solvent radical anions were all in nonpolar liquids. What happens in polar liquids? In some polar liquids (whose molecules have negative $EA_g$) such as water, mono- and poly- atomic alcohols, and ethers, "solvated electron" is observed. In this species, the excess electron density is mainly outside the solvent molecules. The electron resides at the cavity center and interacts as a point-like negative charge with the surrounding solvent dipoles. The greater the dipole the more stable is the "solvated electron". A link between these energetics and the solvent polarity is observed in the bell-like vis or NIR spectra of the "solvated electrons" [1]. The maximum of the band systematically shifts towards the blue as the polarity of the solvent increases. For obvious reasons, these "solvated electrons" are not observed in liquids whose molecules have positive $EA_g$. However, there is no *a priori* reason to believe that a given negative-$EA_g$ polar liquid will yield "solvated electrons" upon ionization. Even if the monomer has negative $EA_g$, the dimer may have positive electron affinity, especially in a liquid solution where the electrostatic field of the solvent (considered as a polarizable dielectric continuum) can stabilize the corresponding dimer anion. For hydroxylated molecules, such as water and aliphatic alcohols, stabilization via the formation of a dimer (or multimer) anion is lacking, because no low-lying unoccupied MO's are readily accessible for the excess electron. By contrast, many organic molecules have readily accessible C and N $2p$ orbitals in their LUMO's; these are the atomic orbitals involved in the formation of C-C bound dimer anions in $CO_2$ and $CS_2$ considered in sections II and III, respectively.

Below, we consider a polar, negative-$EA_g$ liquid - acetonitrile - in which the dimer anion formation (due to the electron accessing low-lying $\pi^*$ orbitals) competes with electron stabilization due to the formation of a polarized solvent cavity [30,46]. The outcome of this competition depends on the solvent temperature, as the dimer anion is in a dynamic equilibrium with a more energetic cavity electron. The latter cannot be regarded as "solvated electron" since the partition between the valence electrons in the solvent molecules and the excess electron at the cavity center is incomplete [30]. We argue that the properties of these two electron states can only be understood when the traditional one-electron approximation is abandoned in favor of many-electron model.

Like water and aliphatic alcohols, gas phase $CH_3CN$ monomer has a large dipole moment (4.3 D) and negative vertical $EA_g$ of -2.84 eV (adiabatic $EA_g$ is +17 meV) [47]. $CH_3CN^-$ is a classical example of a dipole-bound anion, with the electron in a diffuse orbital (> 3 nm) [47]. While neutral dimers, in which the $CH_3CN$ dipoles are coupled in an antiparallel fashion, readily form in vapor and in liquid [48], the dimer anion, $\{CH_3CN\}_2^-$, has not been observed in the gas phase. In the neutral trimer, one of the monomers couples sideways to the antiparallel pair; this molecule binds the electron in the same way as the monomer; the adiabatic electron affinity of this trimer (14-20 meV) is higher than that of the monomer [47]. Higher multimer anions, $\{CH_3CN\}_n^-$, were found only for $n > 12$ [49]. *The stabilization of excess electron in solid and liquid acetonitrile is* *a concerted effect of many solvent molecules.* One would expect that the electron in solid and liquid acetonitrile localizes in the same way as the "solvated/trapped electron" in water and alcohols. This expectation is not borne out.

Solid acetonitrile exists in two crystalline forms, a high-temperature phase, α, and a low-temperature phase, β [50]. When ionized at 77 K, α-acetonitrile yields a dimer radical anion, while β-acetonitrile yields a monomer radical anion [51]. The observed dichotomy follows from the crystal structure: In α-acetonitrile, the dimer anion retains the same reflection plane and inversion center as the symmetric antiparallel pair of $CH_3CN$ molecules [52]. β-Acetonitrile consists of infinite chains of parallel dipoles (no antiparallel pairs are present) and a monomer anion is formed instead [51,52]. EPR experiments and *ab initio* calculations of Williams and co-workers [51] indicate that the dimer radical anion is $C_{2h}$ symmetrical and has the staggered, side-by-side structure shown in Fig. 4a [46,52]. The mechanism for orbital stabilization of bent acetonitrile molecules in the dimer is illustrated in Fig. 4b. The CCN angle is 130º and the distance between cyanide carbons is 0.165 nm. The negative charge and spin are mainly on carbonyl N and methyl C atoms. This structure accounts for the observed EPR parameters and vibronic progressions observed in the charge-resonance band of the dimer radical anion [51,52]. The monomer radical anion in β-acetonitrile is also bent; the CCN angle is close to 131º [52]. In both of these anions, the C-C bond is stretched to 0.153 nm (vs. 0.1443 nm in neutral $CH_3CN$). Photoexcitation of these radical anions (< 650 nm) causes further elongation of the $NC-CH_3$ bonds (due to the promotion of electron into the corresponding C-C antibonding orbital) which leads to their fragmentation to $CH_3$ and $CN^-$ (see Fig. 4b). Except for the vibronic progressions, both radical anions exhibit similar absorption spectra in the visible (see Fig. 3 in ref. [51]). For the dimer radical anion, the absorption band is centered at 530 nm, for the monomer radical anion - at 420 nm [51]. The positions of these bands are in good agreement with *ab initio* calculations [52]. These calculations indicate that no bound-to-bound transitions in the IR are possible, either for the monomer or the dimer radical anion [46,52].

In liquid acetonitrile, there are two radical anions present shortly after the ionization event: anion-1 that absorbs in the NIR (whose band is centered at 1.45 μm) and anion-2 that absorbs in the 400-800 nm region (whose band is centered at 500 nm); see Fig. 5a [46,53]. These two anions are in a rapid dynamic equilibrium (Fig. 5b): as the liquid is cooled, the 1.45 µm band becomes more and the 500 nm band less prominent [53]. From the temperature dependencies of the transient absorption spectra, it was estimated that anion-2 is 0.36 eV more stable than anion-1 [53]. The transformation of anion-1 to anion-2 is rapid at room temperature [40,46,53] but fairly slow at the lower temperature; at -30ºC, it takes 20-50 ns [53].

While it was initially suggested that anion-1 and anion-2 are, respectively, the monomer and the dimer radical anions of acetonitrile [53], more recent work suggests that anion-1 cannot be a monomer anion (which in any case has a different absorption spectrum from anion-1, as explained above) [30,46]. Actually, the absorption spectrum of anion-1 is very similar to that of "solvated electron" in saturated hydrocarbons. This is understandable because the CN dipole has negative charge on the nitrogen. Consequently, if a cavity electron were formed in acetonitrile, this cavity would be lined by *methyl* rather than CN groups [30], i.e., the first solvation shell of the *s* electron would resemble that of the "solvated/trapped electron" in liquid and vitreous alkanes.

The NIR location of the absorption band for a hypothetical cavity electron in acetonitrile makes even more sense if one recalls that there is a linear correlation between the position of the band maximum of a cavity electron in a given polar liquid and the position of the CTTS band maximum for a given halide anion in the same liquid [46,54]. The absorption band of anion-1 fits perfectly on this correlation plot (predicted 1.48 µm [54] vs. the observed 1.40-1.45 µm [46]).

Not only does anion-1 differ from the molecular anions of acetonitrile in its absorption properties, but its dynamic properties are also anomalous. While anion-2 has normal mobility, anion-1 is a high-mobility anion whose room-temperature diffusion coefficient is more than three times higher than that of solute ions and anion-2 [30]. The activation energy for this migration is just 3.2 kJ/mol while the value for normal ions (including anion-2) is 7.6 kJ/mol [30]. Electron-transfer reactions that involve anion-1 proceed with rate constants approaching $10^{11}$ M$^{-1}$ s$^{-1}$ [30,46]. These reactions can be directly observed on a subnanosecond time scale (before the equilibration of the two anions) using ultrafast pump-probe laser spectroscopy. To this end, Kohler and co-workers injected the electron into room temperature liquid acetonitrile using one-photon CTTS excitation of iodide [46]. Both anion-1 and anion-2 were observed within 300 fs after the excitation with a 200 fs, 260 nm pulse, and rapid decay of anion-1 in the presence of $CHCl_3$ was observed [46]. The same experiment gave an estimate of 0.26 ns for the settling of the equilibrium between the two anions.

In the time-resolved photoconductivity experiments carried out at Argonne [30], the anion equilibrium was observed via non-Arrhenius temperature dependencies of anion mobility and rate constants of scavenging by electron acceptors, such as $CCl_4$ [30]. These conductivity experiments clearly demonstrate that the high-mobility anion is anion-1 rather than anion-2, contrary to previous suggestions [55]. To distinguish between the two anions, anion-1 and anion-2 were photoexcited in their respective absorption bands using 1064 nm and 532 nm, 6 ns fwhm laser pulses; this photoexcitation causes anion fragmentation to $CH_3$ and $CN^-$ (with quantum yields of 0.01 and 0.32, respectively) and a decrease in the dc conductivity. Using 532 nm photobleaching of anion-2, the equilibrium fraction of this anion between –20°C and 50°C was determined; knowing this fraction, the mobility and reaction constant for each anion were determined and the equilibrium constant (1.3 at 25°C) and the heat of anion conversion (which is ca. –0.46 eV) were estimated. The photon fluence dependencies of the photobleaching efficiency gave estimates for the anion conversion rate. These measurements suggested a longer time constant of 3 ns (vs. 0.26 ns obtained in ref. [46]) for settling the equilibrium between the two anions at 25°C.

The formation of $CH_3CN^-$ in solid β-acetonitrile is due to its favorable crystal structure [50]. According to x-ray diffraction and NMR data, the short-range structure of liquid acetonitrile is similar to that of crystalline α-acetonitrile, with a pentamer as the basic unit [56]. The prevalent orientation of the acetonitrile molecules in the liquid is the antiparallel pair of the type found in α-acetonitrile. Given that dimerization strongly reduces the energy of the anion [46,51,52], it seems likely that the monomer anion cannot form in liquid acetonitrile, where the "special arrangement" of neighboring molecules needed for the formation of the monomer anion is not possible.

While anion-2 is clearly the dimer radical anion of acetonitrile, identification of anion-1 as a cavity electron requires caution. First, we stress that anion-1 cannot be the monomer anion of acetonitrile. The monomer anion does not absorb in the NIR [30,46,52]. For the monomer anion to occur at all, the neighboring acetonitrile molecules should all be oriented in the same direction, as in β-acetonitrile; otherwise, coupling to a neighboring (antiparallel) molecule reduces the overall energy and causes instant dimer formation. It is difficult to see how such a fortuitous orientation could persist for 0.3-3 ns in a room-temperature liquid. Also, it is not clear why a monomer anion would migrate rapidly. The only migration mechanism possible for this anion would be charge hopping. Assuming that this hopping is between neighboring molecules (separated by 0.4 nm) and the diffusion coefficient is $8.3 \times 10^{-4}$ cm$^2$/s (estimated from the room-temperature mobility of $3.3 \times 10^{-4}$ cm$^2$/Vs [30]), the residence time for the charge on a given molecule is 2 ps. This implies that $10^2$-$10^3$ hops occur prior to the transformation of the monomer anion-1 to anion-2. The lowest bending modes of the CCN fragment of acetonitrile molecules and anions are 300-330 cm$^{-1}$ which is equivalent to 0.1 ps in time units. Thus, though the diffusion is fast, the lifetime of a given "monomer anion" is sufficiently long for the structural relaxation; in other words, this "monomer anion" must be a bent species like $CH_3CN^-$ in β-acetonitrile. Thus, the low-barrier resonant charge transfer needed to explain the high mobility would have to be between a strongly bent anion and a linear neutral molecule. Such a process cannot proceed with a low activation energy, since bending of the neutral molecule and solvation of the resulting anion require much energy. Furthermore, never once in a series of these $10^2$-$10^3$ hops could the two molecules involved in the resonant charge transfer be in the antiparallel orientation, since then anion-1 would couple to the neighboring molecule yielding anion-2. It appears that the monomer anion cannot account for any property of anion-1.

It is more likely that the high mobility anion-1 is a multimer anion in which the charge is spread over several acetonitrile molecules, like the analogous species in nonpolar liquids that were examined in section III. Due to the reduction in the charge on the individual molecules, their bending is less strenuous and the barrier for the migration of the multimer anion is low. Such a multimer anion is actually no different from the "solvated electron" in alkanes (see below), which accounts for the striking similarity between the absorption spectra of anion-1 and "solvated/trapped electrons" in saturated hydrocarbons. We suggest that acetonitrile provides a rare example of a liquid in which the "solvated electron" (multimer anion) coexists with a molecular - dimer - radical anion.

To investigate possible structures of the multimer anion, a $\{CH_3CN\}_3^-$ cluster was modeled using a density functional (B3LYP) method [30]. A 6-31+G** basis set that included polarized *(d,f)* and diffuse functions was used and the $C_{3h}$ symmetry was imposed. A "ghost" hydrogen atom with zero charge was placed at the center of the cluster to provide *s*-functions for the "solvated electron". The polarizable (overlapping spheres) continuum model was implemented in the integral equation formalism. The lowest energy state was a "propeller-like" $^2A'$ state shown in Fig. 6. The CCN angle in the acetonitrile subunits is $178°$ in vacuum and $168°$ in solution (vs. $180°$ in the neutral molecule). This bending is considerably smaller than in the monomer and dimer anions (ca. 130°). The solvated cluster anion is compact: the closest methyl hydrogens are 0.171 nm away from the symmetry center. The C-C bond in the acetonitrile subunits is elongated from 0.144 nm to 0.148 nm, while the C-N bond is changed very slightly. In this structure, the SOMO envelopes the whole cluster anion. The main negative nodes are on methyl carbons, while the main positive nodes are at the center of symmetry, on the in-plane hydrogens, and on carbonyl carbons. This structure may be viewed both as a multimer anion and a "solvated electron: the SOMO has a noticeable *s*-character at the symmetry center (ca. 0.34), though the main spin density is on the methyl carbons. The latter atoms exhibit large hyperfine coupling constants (hfcc) for $^{13}C$: the isotropic hfcc is 6.9 mT; the anisotropy is negligible. Isotropic hfcc for methyl protons are relatively small: 0.19 mT for in-plain hydrogens (the principal values of the dipole tensor are -0.29, -0.15, and 0.44 mT) and -0.086 mT for out-of-plane hydrogens (-0.29, -0.16, and +0.45 mT, respectively). The isotropic hfcc for cyanide $^{13}C$ and $^{14}N$ nuclei are 0.4 mT and 0.36 mT, respectively.

The structure bears strong resemblance to the "trapped electron" in saturated hydrocarbons studied by Kevan and co-workers [2]. The "electron" is "solvated" by methyl groups; the positive charge on these groups is increased due to considerable elongation of C-C bonds. This elongation, as demonstrated by our DFT calculations, is the consequence of large electron density on the skeletal carbon atoms. In the semicontinuum model of Kevan and coworkers [2] this (multielectron) interaction is treated in terms of a "polarizable" C-C bond; our calculation justifies their *ad hoc* approach. The size of the solvation cage, the juxtaposition of methyl groups, and the hfcc tensors for methyl protons compare favorably with those obtained experimentally by Kevan and co-workers for the "trapped electron" in frozen 3-methylpentane [2]. Therefore, it is reasonable that the multimer $\{CH_3CN\}_n^-$ anion absorbs much like the "solvated/trapped electron" in alkanes.

While a first-principle calculation for a larger cluster is impractical, it is possible to make an educated guess as to what happens to the anion when the cluster size increases. The "propeller" structure obtained for the $\{CH_3CN\}_3^-$ anion is similar (save for the elongated C-C bonds) to that of the $\{CH_3CN\}_n X^-$ (X=I, Br) cluster for *n=3* [57]. One may expect that this trend will pertain to larger size clusters. When the halide anion is solvated by less than seven acetonitrile molecules, the core anion is a "star" structure with radial $CH_3CN$ dipoles looking away from the halide anion [57]. For *n>9-12*, the molecules in the first solvation shell couple in an antiparallel fashion to the molecules in the second solvation shell, so that some molecules in the first solvation shell are oriented tangentially rather than radially [57]. Perhaps, small $\{CH_3CN\}_n^-$ anions (n≤6) are also star-shaped. Due to the further spread of the electron density in such clusters, the *s*-character of the SOMO increases while the CCN bending and C-C bond elongation decreases: such an anion would be more like a "solvated electron".

To conclude this section, acetonitrile is an example of a polar liquid in which stabilization of the excess electron via the formation of a dimer anion is favored energetically over the formation of a cavity electron, despite the fact that the molecule has one of the largest dipole moments and very negative $EA_g$. The cavity electron still occurs in this liquid as a metastable state at the high-temperature. This state cannot be truly regarded as solvated electron since the electron density is shared both by the solvent molecules and interstitial sites; the excess electron is not separable from the valence electrons of the solvent. A similar situation exists for "trapped electrons" in vitreous hydrocarbons. These species should be regarded as multimer anions with flexible geometry and extensive delocalization of the charge. These "solvated electrons" are just variants of multimer radical anions that occur in many liquids, both polar and nonpolar, including the several examples examined above.

**V. Solvent radical cations in liquid cycloalkanes.**

At first glance, it may appear that extensive delocalization and/or rapid charge hopping should not occur for solvent radical *cations* because the valence "hole" is more strongly associated with the molecule than the excess electron. We have already seen that such expectations are not supported for solvent anions, were the delocalization and degenerate electron exchange occur for liquid solvents whose molecules differ by more than 2 eV in their electron affinity. The same applies to the solvent holes: the fact that a given molecule forms a well-defined radical cation when this molecule is isolated in an inert matrix does not mean that the same species is formed in a liquid where all molecules are alike. The last few examples discussed in this paper are high-mobility solvent holes in cycloalkanes: cyclohexane, methylcyclohexane, and decahydronaphthalenes (decalins) [58,59,60].

We forewarn the reader that the formation of high-mobility holes is not peculiar to these four cycloalkanes: for instance, cyclooctane [61], squalane [62,63,64], and $CCl_4$ [65] also yield such holes. However, in these other liquids, the holes are unstable and, consequently, more difficult to study (the lifetimes are 5-20 ns vs. 1-3 μs). This explains why convincing demonstrations for the occurrence of high mobility holes are slow to come. E.g., squalane (by virtue of its high viscosity) has been frequently used in the studies on fluorescence and magnetic and spin effects in pulse radiolysis. Despite these many studies, only recently has it been recognized that its short-lived hole (with lifetime < 20 ns) has abnormally high diffusion and reaction rates [62]. Shortly after this fact was established using transient absorption spectroscopy, subsequent studies confirmed the hopping mechanism, as fast diffusion with degenerate electron exchange and high-rate scavenging reactions of the squalane holes were observed using time-resolved ODMR [62], magnetic level-crossing, and quantum beat spectroscopies [63]. Rapid scavenging reactions of the squalane hole were also found to account for the anomalies in the magnetic field effect observed for delayed fluorescence in the VUV excitation of squalane [64]. Basically, in such systems, one needs to know where to look; once the property is established, it can be demonstrated in several ways, using different techniques.

In cyclooctane, high-mobility solvent holes were observed using time-dependent electric-field-modulated delayed fluorescence [61] and by observation of rapid scavenging of cyclooctane holes by aromatic solutes in the initial stage of radiolysis. Recently there has been a suggestion of the presence of such holes in cyclopentane and cycloheptane [61]; their natural lifetimes must be < 5 ns. Faster-than-normal scavenging of short-lived isooctane holes by diphenylsulfide and biphenyl was observed using quantum beat and transient absorption spectroscopies [66]. A controversy exists as to the presence of high-mobility holes in liquid $CCl_4$ [65].

These disparate findings hint that there may be many examples of rapidly migrating (delocalized) solvent holes in molecular fluids: the known systems are few because it is difficult to establish these properties for short-lived species. As the time resolution improves, more examples might follow. In most saturated hydrocarbons, fragmentation and proton transfer limit the lifetime of the solvent hole to several nanoseconds (or less) [58] and, therefore, little is known about their dynamics. On the other hand, the most studied alkane liquids, paraffins, do not seem to yield high-mobility solvent cations [67]. This is due to the fact that many conformers coexist in these liquids, some of which have higher ionization potential than others. Variations in the binding energy of the hole stall its rapid hopping since thermal activation is needed to detrap the hole from the low-IP conformers. That conformation dynamics and isomerism play an important role in the charge hopping is supported by many observations (note, for example, that high mobility solvent *anions* are known to occur only in liquids whose molecules are rigid). As for the paraffins, while no rapid hole hopping is observed in *liquid* alkanes, in low-temperature crystals (where all molecules have the same extended conformation) this exchange is very fast and can be readily observed by means of time-resolved and/or cw ODMR [68]. Ironically, in these *n*-alkane crystals, the hole migrates much faster than the hole in frozen *cyclo*alkanes, because the latter solids exhibit more structural disorder (due to formation of plastic crystal and glass phases) detrimental to hole hopping; thus, the situation is exactly opposite to that in a liquid. The recent magnetic level-crossing spectroscopy study of Borovkov et al. places an upper estimate of just $10^8$ M$^{-1}$ s$^{-1}$ for degenerate electron exchange between *n*-nonane$^+$ and the parent alkane molecule in room-temperature solution [68].

The reader may notice that only saturated hydrocarbons (with a possible exception of CCl$_4$) have been observed to yield rapidly migrating solvent holes. As mentioned above, part of this bias is explained by the fact that the holes are usually short-lived, so their dynamic properties are difficult to study. However, in many liquids (such as aromatic hydrocarbons and sc CO$_2$) the solvent holes are relatively stable, yet no rapid hole hopping is observed. In such liquids, the solvent hole has a well-defined dimer cation core with strong binding between the two halves (in the first place, it is this dimerization that causes the hole stability). For example, solvent holes in aromatic liquids are sandwich dimer cations with overlapping π systems [43]; in sc CO$_2$, the solvent cation is an O-O bound molecular dimer [19], etc. This strong dimerization is detrimental to charge delocalization and rapid hopping. High temperature is needed to overcome this hindrance; perhaps high-mobility holes more readily occur in hot (e.g., supercritical) liquids. In many room-temperature liquids, a catch-22 situation occurs: for the solvent radical cations to be stable towards fragmentation and proton transfer, these holes must dimerize. The dimer radical cations are long lived and can readily be studied, however, they have ordinary dynamic properties. The holes that do not dimerize might have interesting dynamic properties but they are unstable and, therefore, difficult to study. As a result, one is limited to the studies of the few solvent holes that do not dimerize and yet are long-lived.

In cycloalkanes, proton transfer is weakly endothermic, conformational dynamics is slow, dimerization is not favored, and the high mobility solvent holes can be readily observed [60]. Ionization of cyclohexane, methylcyclohexane, *trans*-decalin and *cis*-decalin produces cations whose mobilities are 5-to-25 times greater than the mobilities of normally-diffusing molecular ions and (in some cases) thermalized electrons in these liquids [58,59,60]. Long lifetime and high mobility makes it possible to study the reactions of these holes using time-resolved microwave and dc conductivity, an option that does not exist for other saturated hydrocarbons. The activation energies for the hole mobility range from -(3±1) kJ/mol for *trans*-decalin and cyclohexane to +(7-8) kJ/mol for methylcyclohexane and *cis*-decalin [58]. Methylcyclohexane has the largest temperature interval where it is liquid at atmospheric pressure and exhibits a single activation energy of hopping (7.8 kJ/mol) between 133 and 360 K [69]. The activation energies for the highest-rate scavenging reactions of the cycloalkane holes range from 4 kJ/mol to 9 kJ/mol [58]. All these activation energies are small, suggesting low barrier for resonant charge transfer.

Dynamic and chemical properties of the cycloalkane holes have been reviewed [58,59], and we refer the reader to these publications for more detail. Below, we briefly summarize the main findings. Although the cycloalkane holes are paramagnetic species, these holes cannot be observed by magnetic resonance techniques, whether in neat cycloalkanes or in dilute solutions in high-IP liquids. Only recently has it been understood that rapid spin-lattice $(T_1)$ relaxation in the high-symmetry cycloalkane radical cations precludes their detection using ODMR [70]. This relaxation is caused by dynamic averaging between the nearly degenerate ground and excited states of the radical cations; this degeneracy results from the Jahn-Teller distortion. For example, *trans*-decalin cation isolated in room-temperature cyclohexane has $T_1 < 7$ ns [70]. Since it takes several tens of nanoseconds to flip the electron spin for detection, radical cations of these cycloalkanes cannot be detected by ODMR.

In the early studies, the cycloalkane holes were viewed as molecular radical cations that undergo rapid resonant charge transfer. At any given time, the positive charge was assumed to reside on a single solvent molecule and, once in 0.5-2 ps, to hop to a neighboring molecule. The low activation energy was explained by the similarity between the shapes of cycloalkane molecules and their radical cations [60].

This model is consistent with many observations. Dilution of cycloalkanes with high-IP alkanes (or higher-IP cycloalkanes) results in a decrease in the hole mobility that correlates with the mole fraction of the cycloalkane in the mixture: the hopping rate decreases when the density of the like molecules decreases. The occurrence of resonant charge hopping is firmly established experimentally. Charge transfer between $c$-$C_6D_{12}^+$ and $c$-$C_6H_{12}$ was observed in the gas phase, where it proceeds at 1/3 of the collision rate [71]. The hopping was also observed for radical cations and molecules of *cis*- and *trans*-decalins in dilute cyclohexane solutions (where it proceeds with a diffusion-controlled rate) [70]. In low-temperature solid hydrocarbons (4-30 K), hole hopping was observed by ODMR [68]. At higher temperatures, the spectral diffusion caused by the hopping causes the ODMR spectrum to collapse to a single narrow line observed using magnetic level-crossing and quantum beat spectroscopies [64,65].

On the other hand, matrix-isolation EPR and *ab initio* calculations suggest that neutral cycloalkanes and their cations have rather different geometries. In *cis-* and *trans-*decalins the bridging bond elongates from 0.153-0.156 nm in the neutral molecule to 0.19-0.21 nm in the radical cation [73]. Upon charging, the molecules undergo considerable structural relaxation, losing 0.5-0.7 eV [73]. If the electron transfer were a single-step process, it would require an activation energy of 1-2 eV. What then makes the resonant charge transfer possible? In the gas phase, the electron exchange proceeds through the formation of a collision complex in which the charge is shared by both of the cycloalkane moieties [71]. This sharing considerably reduces the barriers for the structural relaxation. It may be assumed that in liquid cycloalkanes the charge is shared between several solvent molecules (analogous to the situation for solvent anions) and this sharing further reduces the hopping barrier.

The sharing of charge causes delocalization of the hole. The best evidence for the delocalization of cycloalkane holes was provided by large scavenging radii (> 2 nm) in cold methylcyclohexane [69] and by hole dynamics in cyclohexane-methylcyclohexane mixtures [74]. While the addition of less than 5-10 vol % of methylcyclohexane to cyclohexane reduces both the dc conductivity signal and its decay rate, further addition of methylcyclohexane yields little change in the conductivity signal and kinetics. The initial reduction is accounted for by rapid reversible trapping of cyclohexane holes by methylcyclohexane [74]. At higher concentration of methylcyclohexane, the equilibrium fraction of the cyclohexane holes becomes very low and the conductivity should decrease. Experimentally, the migration of methylcyclohexane hole in 5 vol % methylcyclohexane solution is as rapid as that of the solvent holes in neat methylcyclohexane. When the methylcyclohexane is diluted by *n*-hexane instead of the cyclohexane, the conductivity signal decreases proportionally to the fraction of *n*-hexane. These results suggest that the methylcyclohexane holes are coupled to the cyclohexane solvent (the difference in the liquid IPs is < 0.11 eV [74]). This coupling makes the charge migration of methylcyclohexane holes in cyclohexane as efficient as in neat methylcyclohexane. From the critical concentration of methylcyclohexane, the delocalization radius was estimated as 1 nm, or 4 to 5 molecular diameters [74]. Thus, the degree of charge delocalization in cyclohexane (for the hole) and hexafluorobenzene (for the electron) are comparable.

We turn to the chemical behavior of cycloalkane holes. Several classes of reactions were observed for these holes: (i) fast irreversible electron-transfer reactions with solutes that have low adiabatic IPs (ionization potentials) and vertical IPs (such as polycyclic aromatic molecules); (ii) slow reversible electron-transfer reactions with solutes that have low adiabatic and high vertical IPs; (iii) fast proton-transfer reactions; (iv) slow proton-transfer reactions that occur through the formation of metastable complexes; and (v) very slow reactions with high-IP, low-PA (proton affinity) solutes.

*Class (i) reactions* were observed in all four cycloalkanes. The highest rate constants were observed for reactions of cyclohexane hole with low-IP aromatic solutes, $(3-4.5) \times 10^{11}$ $M^{-1}$ $s^{-1}$ at $25^{\circ}C$ [75]. In these irreversible reactions, a solute radical cation is generated. *Class (ii) reactions* were observed for reactants 1,1-dimethylcyclopentane, *trans*-1,2-dimethylcyclopentane, and 2,3-dimethyl-pentane in cyclohexane [74], *trans*-decalin, bicyclohexyl, and *iso*-propylcyclohexane in methylcyclohexane [69], and benzene in *cis*- and *trans*-decalins [76] (Fig. 7). In these class (ii) reactions, biexponential scavenging kinetics of the solvent hole results due to the dynamic equilibrium between the solvent hole and the corresponding solute cation (in the latter case, the kinetics are complicated by the subsequent dimerization of the benzene cation, Fig. 7). For methylcyclohexane in cyclohexane, the equilibrium is reached so rapidly that the decay kinetics are single exponential at any temperature. Similar equilibria exist for high-mobility holes in mixtures of *cis*- and *trans*-decalins.

The rate constants of the forward class (ii) reactions are much slower than those of the class (i) reactions, though some of electron donors have comparably low adiabatic $IP_g$'s. These rate constants do not correlate with the free energies of hole scavenging reactions obtained from the temperature dependencies of equilibria parameters [74]. An explanation proposed was that the rate constants are controlled by the height of the activation barrier determined by the difference in the *vertical* IP of the solute and the adiabatic IP of the solvent (section II) [74]. A similar mechanism accounts for the chemical behavior of the high mobility solvent radical anion in sc $CO_2$ (*vide supra*) [18].

*Class (iii) reactions* include proton-transfer reactions of solvent holes in cyclohexane and methylcyclohexane [71,74,75]. The corresponding rate constants are 10-30% of the fastest class (i) reactions. *Class (iv) reactions* include proton-transfer reactions in *trans*-decalin and *cis-trans* decalin mixtures [77]. Proton transfer from the decalin hole to aliphatic alcohol results in the formation of a C-centered decalyl radical. The proton affinity of this radical is comparable to that of a single alcohol molecule. However, it is less than the proton affinity of an alcohol dimer. Consequently, a complex of the radical cation and alcohol monomer is relatively stable towards proton transfer; when such a complex encounters a second alcohol molecule, the radical cation rapidly deprotonates. Metastable complexes with natural lifetimes between 24 ns (2-propanol) and 90 ns (*tert*-butanol) were observed in liquid *cis-* and *trans-* decalins at $25^{\circ}C$ [77]. The rate of the complexation is 1/2 of that for class (i) reactions; the overall decay rate is limited by slow proton transfer in the 1:1 complex. The rate constant of unimolecular decay is $(5-10) \times 10^6 \text{ s}^{-1}$; for primary alcohols bimolecular decay via proton transfer to the alcohol dimer prevails. Only for secondary and ternary alcohols is the equilibrium reached sufficiently slowly that it can be observed at $25^{\circ}C$ on a time scale of > 10 ns. There is a striking similarity between the formation of alcohol complexes with the solvent holes (in decalins) and solvent anions (in sc $CO_2$).

A detailed analysis of the thermodynamics and energetics of the complexation reactions is given in ref. [77]. The forward reaction has near-zero activation energy, whereas the proton transfer within the complex is thermally-activated (20-25 kJ/mol). The stability of the complex increases with the carbon number of the alcohol; the standard heat of the complexation decreases in the opposite direction (from -39 kJ/mol for ethanol to -25 kJ/mol for *tert*-butanol). Complexes of *cis*-decalin$^+$ are more stable than complexes of *trans*-decalin$^+$ since for the former, the standard reaction entropy is 35 J mol$^{-1}$ K$^{-1}$ more positive. The decrease in the entropy is small for both decalins (> - 80 J mol$^{-1}$ K$^{-1}$) and approaches zero for higher alcohols. Similarly small changes in the standard entropy were observed for class (ii) reactions of the methylcyclohexane hole [69]. Since the molecular complex formation can only reduce the degrees of freedom, to account for the small change in the entropy there must be an increase in the solvent disorder. This is consistent with a hole ordering solvent molecules around itself. When the positive charge is compensated, the solvent becomes disordered, and the reaction entropy increases. The same effect is expected to occur for all solvent radical ions considered in this paper.

Extremely slow *class (iv)* reactions were observed for scavenging of (a) cyclohexane hole by cyclopropane [60] and (b) cyclohexane and decalins holes by $O_2$ [75]. H atom transfer from the hole to $O_2$, and $H_2^-$ transfer from cyclopropane to the hole were suggested as the possible reaction mechanisms.

In conclusion, the behavior of high-mobility solvent anions and cations is similar. Both occur only in liquids whose molecules have rigid bodies and exhibit little or no conformational dynamics. Both do not occur in liquids where solvent radical ions have a strong tendency to form dimers with neutral solvent molecules. Both migrate by rapid hopping - sometimes over the entire liquid range of the solvent - and involve charge delocalization over several molecules. The activation energies and the degree of delocalization are roughly the same. Delocalization is required for the hopping to be rapid because it reduces geometric adjustment to charge placement and thereby decreases the activation barrier for charge transfer. Both species rapidly react with electron donors/acceptors with rate constants that are determined only by vertical $IP_g$ or $EA_g$ of the solute. In nonpolar liquids, both species display a strong tendency to form metastable complexes with polar molecules, such as alcohols and nitriles, in which the charge is electrostatically bound to the solute dipole. With respect to this propensity, the high-mobility ions are similar to "solvated electrons" in saturated hydrocarbons. Even in polar solvents, solvent anions (e.g., the dimer anion in acetonitrile) are protonated only after formation of a complex with the alcohol monomer; the transfer occurs when a second alcohol molecule encounters the complex [30].

## VI. Concluding remarks.

The take-home lesson of this paper is that there are many ways in which a charge can be localized in a molecular system, and quite a few liquids localize electrons and holes in ways that defy easy classification. One does not need to look far for such "exotic" systems; ordinary solvents will do. In liquid acetonitrile [30,46], a high-energy electron state, a cavity electron, coexists in a dynamic equilibrium with a low-energy state, a dimer radical anion. In liquid benzene [45], the negative charge can migrate both by thermal emission into the CB and by degenerate electron hopping, depending on the pressure. Actually, most liquids seem to exhibit unique charge dynamic properties; there are few general rules.

In the previous four sections, several solvent radical ions that cannot be classified as molecular ions ("a charge on a solvent molecule") were examined. These delocalized, multimer radical ions are intermediate between the molecular ions and "cavity electrons", thereby bridging the two extremes of electron (or hole) localization in a molecular liquid. While "solvated electrons" appear only in negative-$EA_g$ liquids, delocalized solvent anions appear both in positive and negative-$EA_g$ liquids. Actually, from the structural standpoint, "trapped electrons" in low-temperature alkane and ether glasses [2] are closer to the multimer anions since their stabilization requires a degree of polarization in the molecules that is incompatible with the premises of one-electron models.

How general is the formation of multimer solvent ions? We reiterate the argument made in section III that very few systems apart from water, alcohols, saturated hydrocarbons, and ethers have been studied by pulse radiolysis and laser photolysis, and for most of these liquids the ionic species observed were not *primary* ions. The incidence of high-mobility *primary* ions among neat organic liquids is actually high. It should also be kept in mind that if a given liquid does not yield high-mobility solvent ions under normal conditions, this does not necessarily hold for other conditions. High-mobility solvent anions in sc $CO_2$ occur only in the supercritical phase [18,20]; in the cold liquid, the binding of the dimer anion core is too strong for the rapid charge hopping to occur. High-mobility solvent anions in benzene [45] are observed only under high pressure conditions, etc. Furthermore, as discussed above, many organic liquids yield solvent ions that are short-lived ( < 10 ns), and their dynamic properties cannot be studied using existing pulse radiolysis techniques. Finally, only in a small subset of liquids (whose molecules have rigid bodies and whose ions do not dimerize), can the delocalization of the excess charge be observed through faster-than-Brownian-diffusion hopping.

The authors believe that the formation of "peculiar" solvent ions is common; however, only in a handful of cases can one clearly demonstrate that such ions are formed. Far from being exotic species, these ions may constitute the rule, whereas the textbook species, "solvated electrons" and molecular ions, could be rare exceptions. That these exceptions loom large in the collective mind of chemical physicists is due to the fact that aqueous solutions surround us in everyday life, and most radiation chemistry studies have been carried out on aqueous solutions at $25^{\circ}$C. Overcoming this anthropocentric bias, by expanding the range of physical conditions and the number of systems studied, might rejuvenate the 21st century radiation chemistry.

**Acknowledgment.**

The preparation of this review was supported under contract No. W-31-109-ENG-38 with US-DOE Office of Basic Energy Sciences, Division of Chemical Sciences.

1/12/04

**Figure captions.**

Fig. 1

(a) The structure of $D_{2d}$ symmetric $(CO_2)_2^-$ dimer radical anion. (b) Visualization of rapid resonant charge hopping in sc $CO_2$. The hopping barrier is 0.46 eV; the residence time of the charge on a given (dimer) molecule is 0.6 to 4 ps, depending on the solvent temperature [18]. (c) Reduced density ($\rho_r$) dependence of solvent anion mobility $\mu(-)$ for four temperatures. The activation energy does not depend on density; the mobility exponentially increases with the solvent density.

Fig. 2

(a) When metastable quasifree electron, $e_{qf}^-$ (with mobility > 10 cm$^2$/Vs) is trapped by dense sc $CO_2$ solvent, a high-mobility radical anion, $(CO_2)_n^-$, with $\mu(-)$ of 10$^{-2}$ cm$^2$/Vs is formed. The binding energy of the electron is 1.6-1.8 eV; the electron can be detrapped by absorption of a photon with energy > 1.76 eV. (b) Electron photodetachment (empty symbols) and photoabsorption (filled circles) spectra of solvent radical anion in sc $CO_2$. The arrow points to the onset of the photodetachment band. (c) Photoinduced electron detachment in sc $CO_2$ observed by dc photoconductivity. The initial (clipped) narrow peak is the prompt conductivity signal from free electrons; the time profile of this signal follows the shape of the 248 nm excitation laser pulse. The arrows indicate the delay times at which a second (532 nm) laser pulse was fired. 532 nm photoexcitation detraps the electrons that subsequently trap within the duration of the 532 nm laser pulse (sharp "spikes").

Fig. 3

Reversible electron transfer reaction of the solvent radical anion in sc $CO_2$ with dioxygen. The resulting $O_2^-$ anion rapidly forms O-O bound complex with a solvent molecule, yielding a stable radical anion, $CO_4^-$. (a) 532 nm photoexcitation can detach an electron from the solvent anion (Fig. 1a). $CO_4^-$ is a 0.44 eV deeper trap and the electron cannot be detached from it. By determining the magnitude of the "spike" from quasifree electrons (generated by 532 nm laser excitation of the pholysate) as a function of time one can obtain the decay kinetics of the solvent anion. (b) Experimental realization of this concept for a sc $CO_2$ solution containing 120 μm of $O_2$. The conductivity signal ($\sigma$) shown on the double logarithmic scale shows a gradual transformation of the solvent anion to $CO_4^-$. The ratio $\Delta\sigma/\sigma$ of the 532 nm laser induced conductivity signal to the conductivity tracks the concentration of the solvent anion in a reaction mixture that contains both $(CO_2)_n^-$ and $CO_4^-$. After the first 10 μs, this fraction persists at a small value; this represents the settling of the equilibrium between the solvent anion and $CO_4^-$.

Fig. 4

(a) Singly occupied molecular orbital (SOMO) of dimer radical anion of acetonitrile (from a density functional calculation); (b) A scheme for the formation of the SOMO and 1/12/04the doubly occupied subjacent orbital from $\pi$ and $\pi^*$ orbitals of neutral acetonitrile molecules.

Fig. 5

(a) Typical end-of-pulse absorption spectra obtained in pulse radiolysis of room-temperature liquid acetonitrile (7 ns fwhm pulse of 20 MeV electrons). The 500 nm peak is from anion-2 (dimer radical anion); the 1450 nm peak is from anion-1 (cavity electron). (b) Energy diagram and sketches of anion-1 and anion-2 (see the text).

Fig. 6.

Singly occupied molecular orbital (SOMO) of a propeller like trimer radical anion of acetonitrile obtained using density functional theory. The structure was "immersed" in a polarizable dielectric continuum with the properties of liquid acetonitrile. Several isodensity surfaces are shown. The SOMO has a diffuse halo that envelops the whole cluster; within this halo, there is a more compact kernel that has nodes at the cavity center and on the molecules.

Fig. 7

A typical reversible reaction of a cycloalkane hole. (a) *Trans*-decalin hole rapidly reacts with benzene transferring positive charge to the solute. The reverse charge transfer reaction is relatively slow (the free energy decreases by 200 meV), and the lifetime of benzene monomer is ca. 7 ns. This lifetime is further shortened by dimerization of the monomer; this dimerization shifts the equilibrium to the right side. The charge transfer competes with slow, with the formation of decalyl radical and benzonium carbocation by an irreversible proton transfer. (b) The populations of solute monomer and dimer cations (e.g., benzene cations) can be tracked using a "hole injection" technique analogous to the electron photodetachment technique discussed in the caption to Fig. 3c. Photoexcited solute radical cations oxidize the solvent and their dynamics can be followed through the observation of increase in conductivity due to generation of the high-mobility solvent hole; this increase is analyzed as a function of the delay time of the excitation pulse [76].

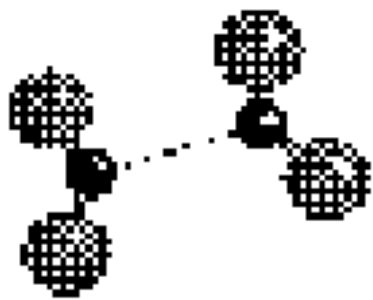
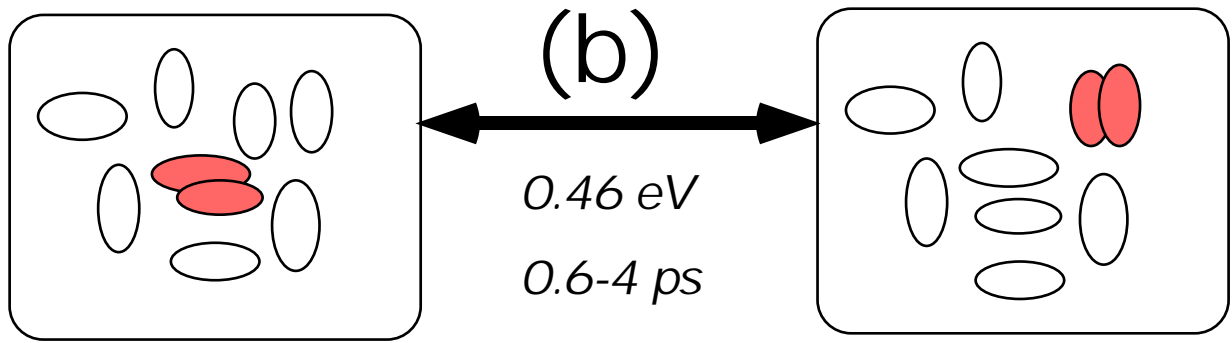
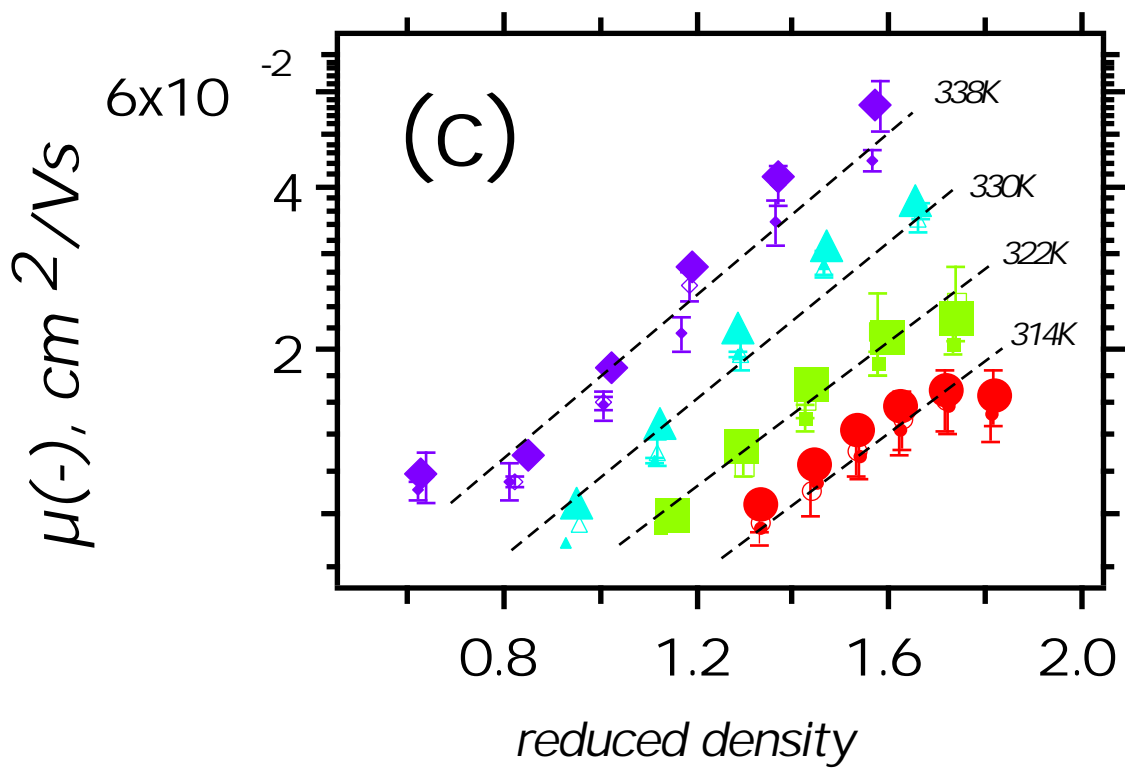

Figure 1

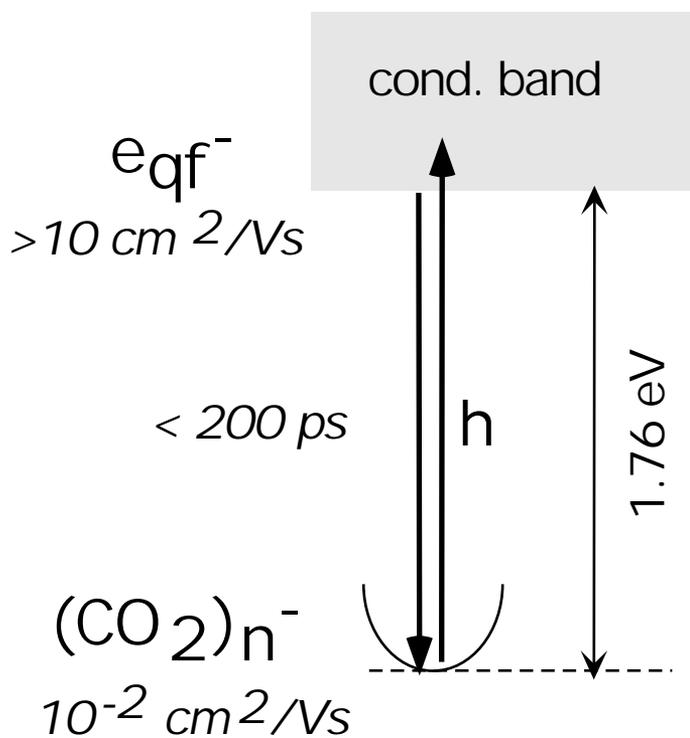
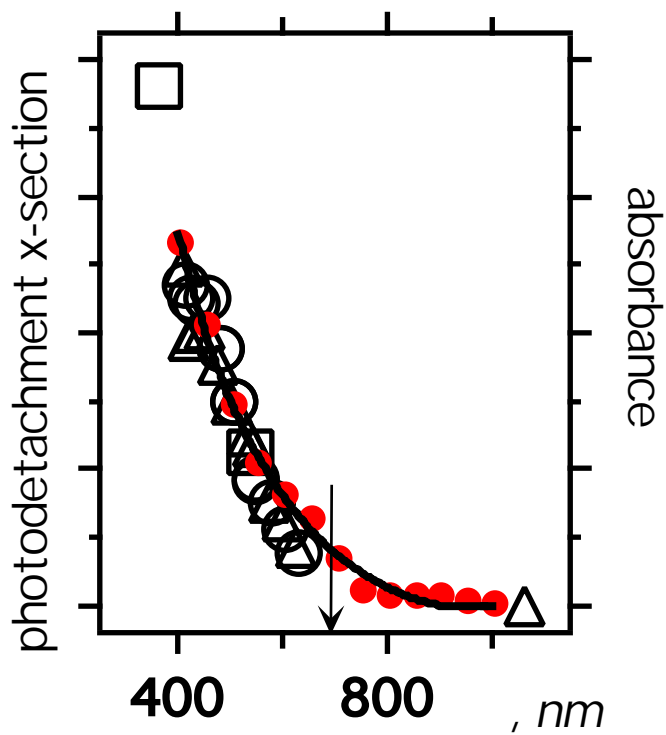
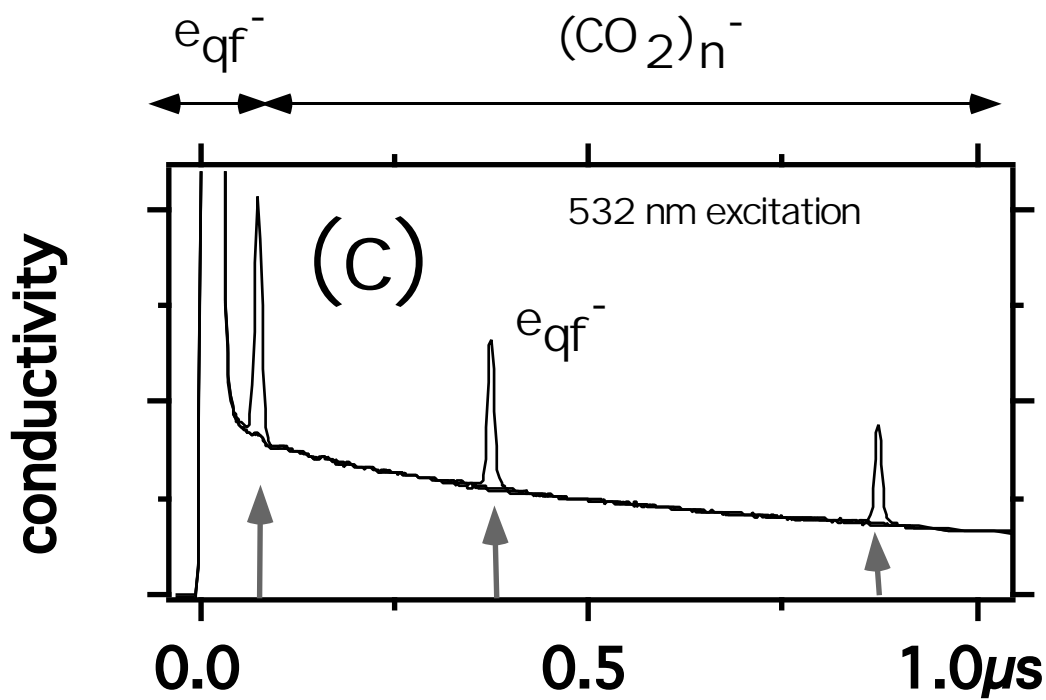

Figure 2

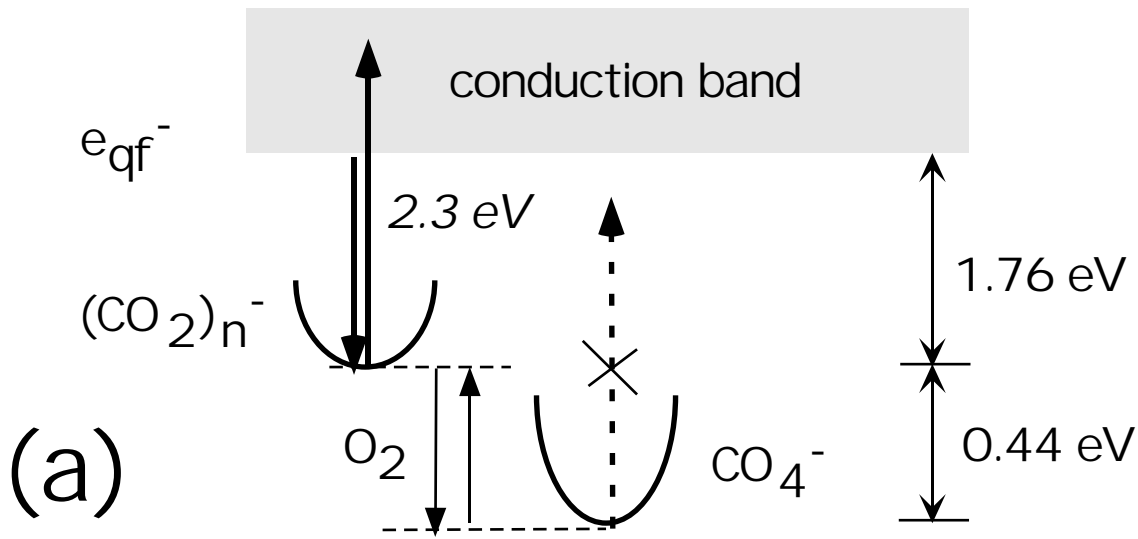

$$(CO_2)_n^- + O_2 \rightleftharpoons (n-1)\,CO_2 + CO_4^-$$

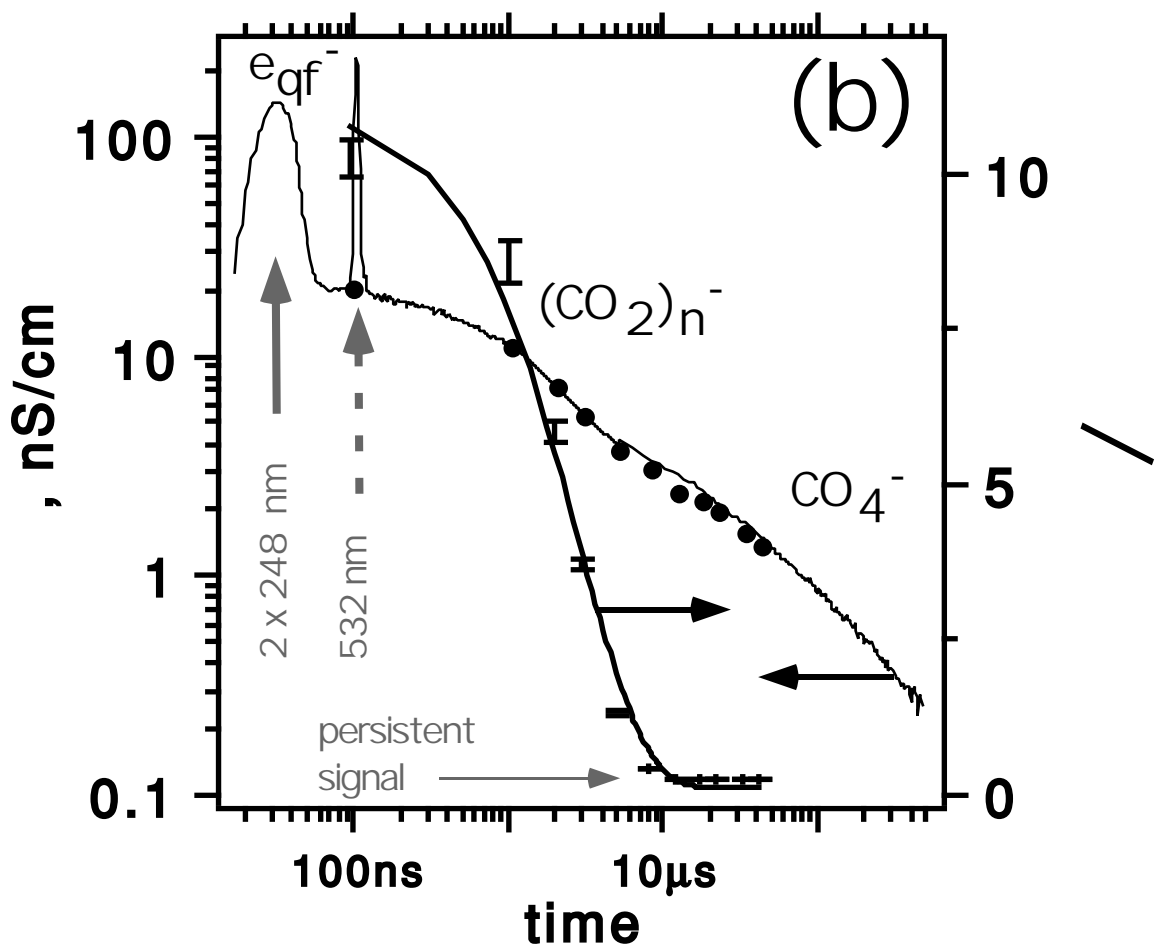

Figure 3

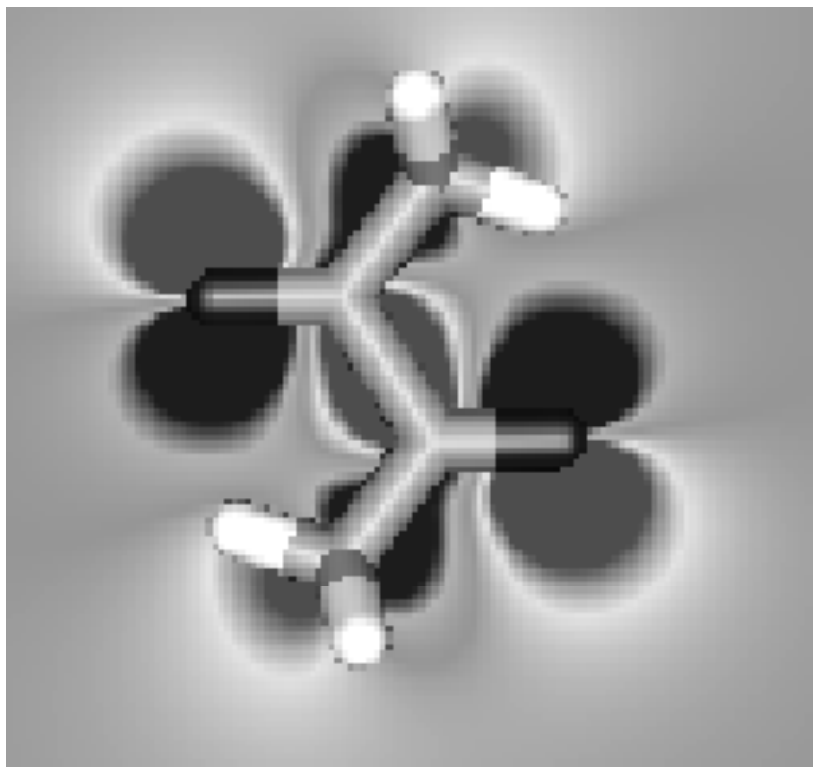

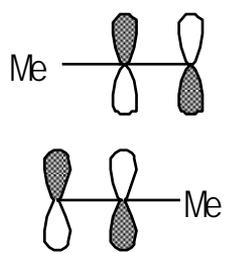
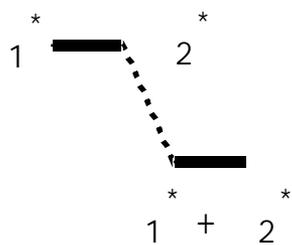
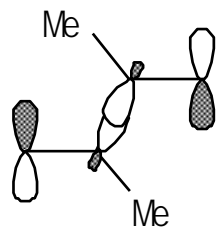
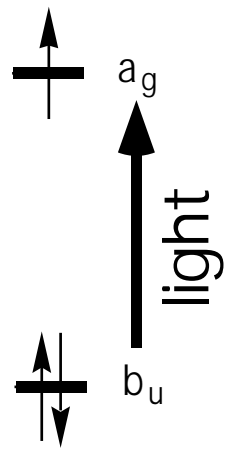

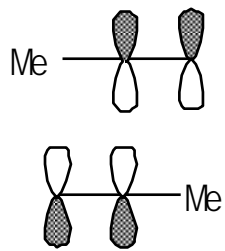
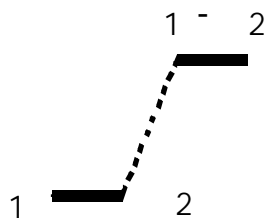
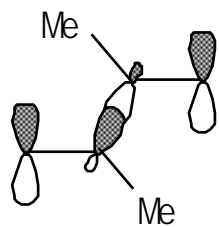

**Figure 4**

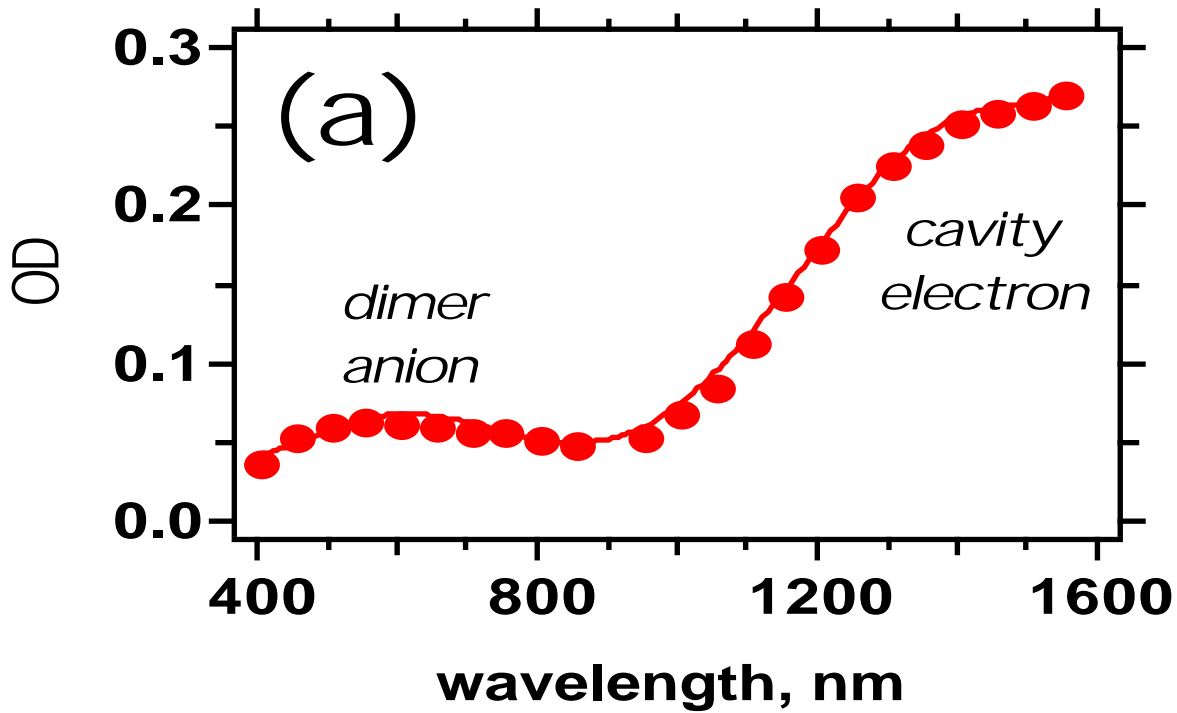
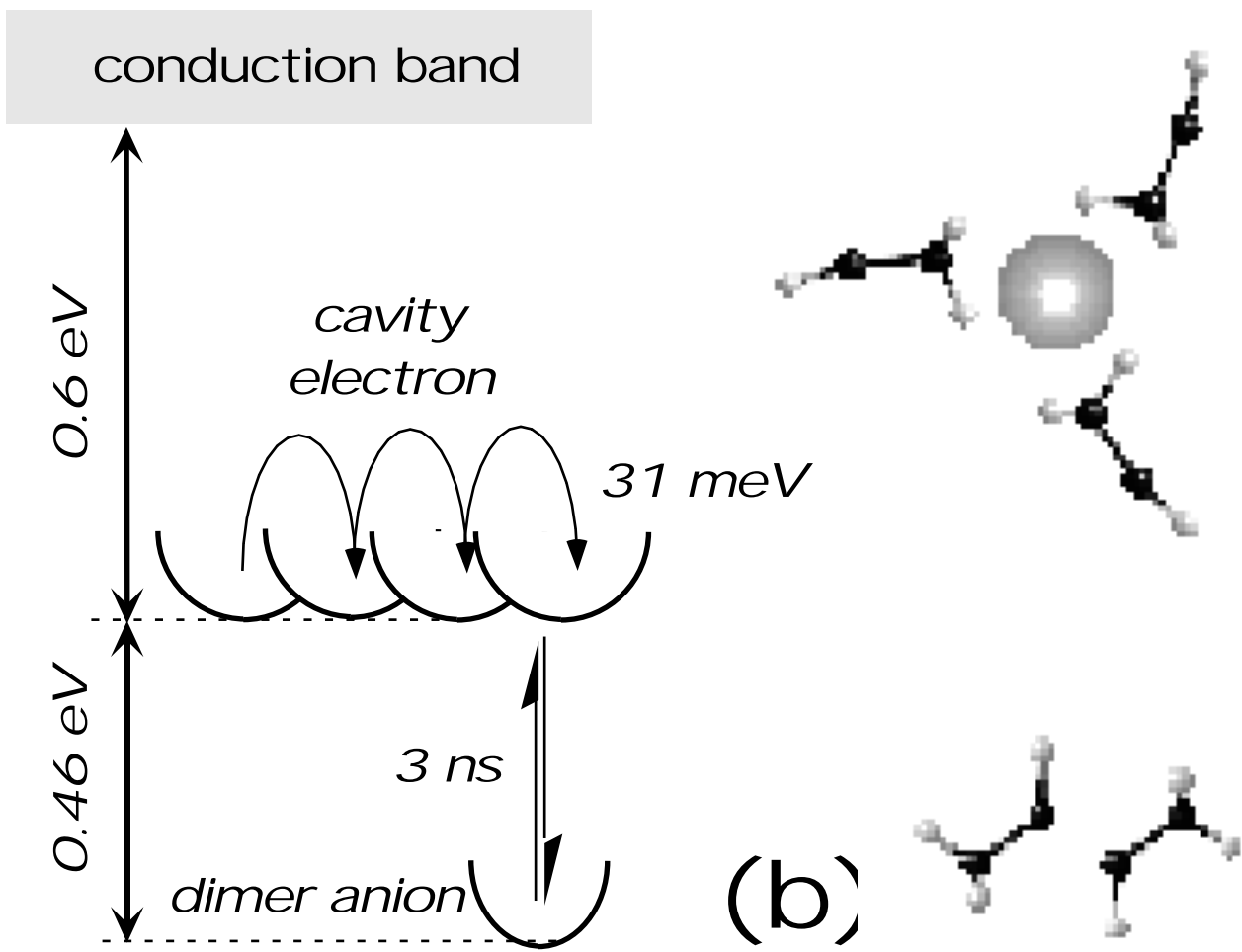

Figure 5

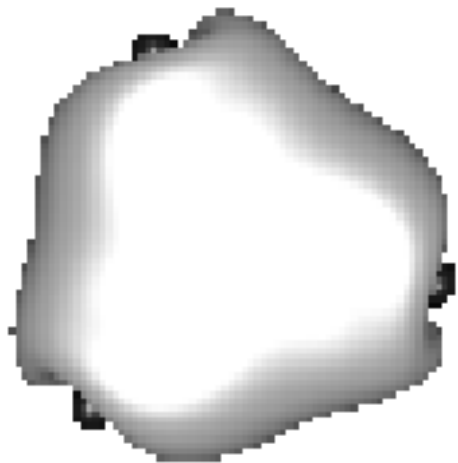
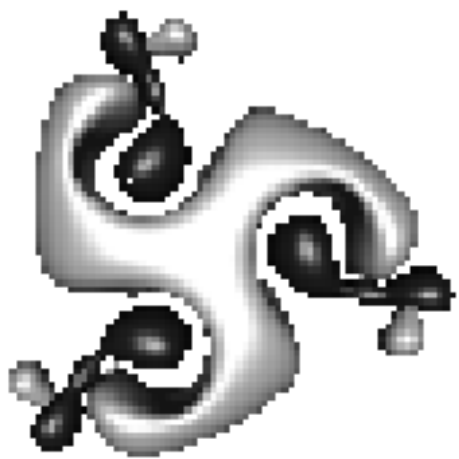
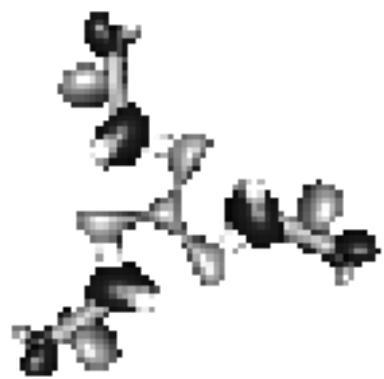

**Figure 6**

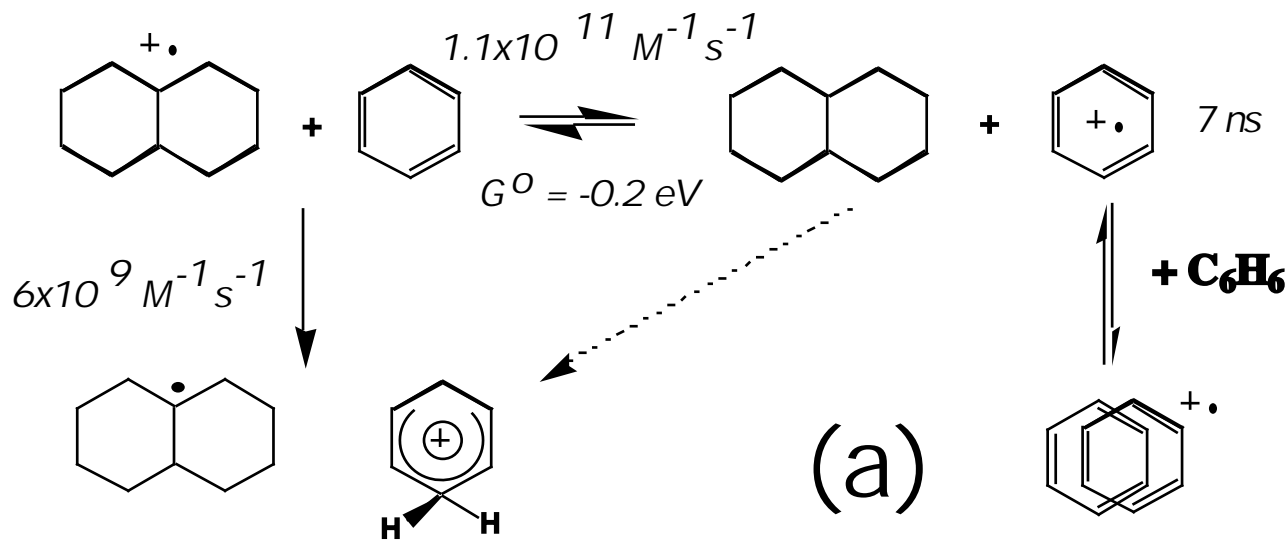

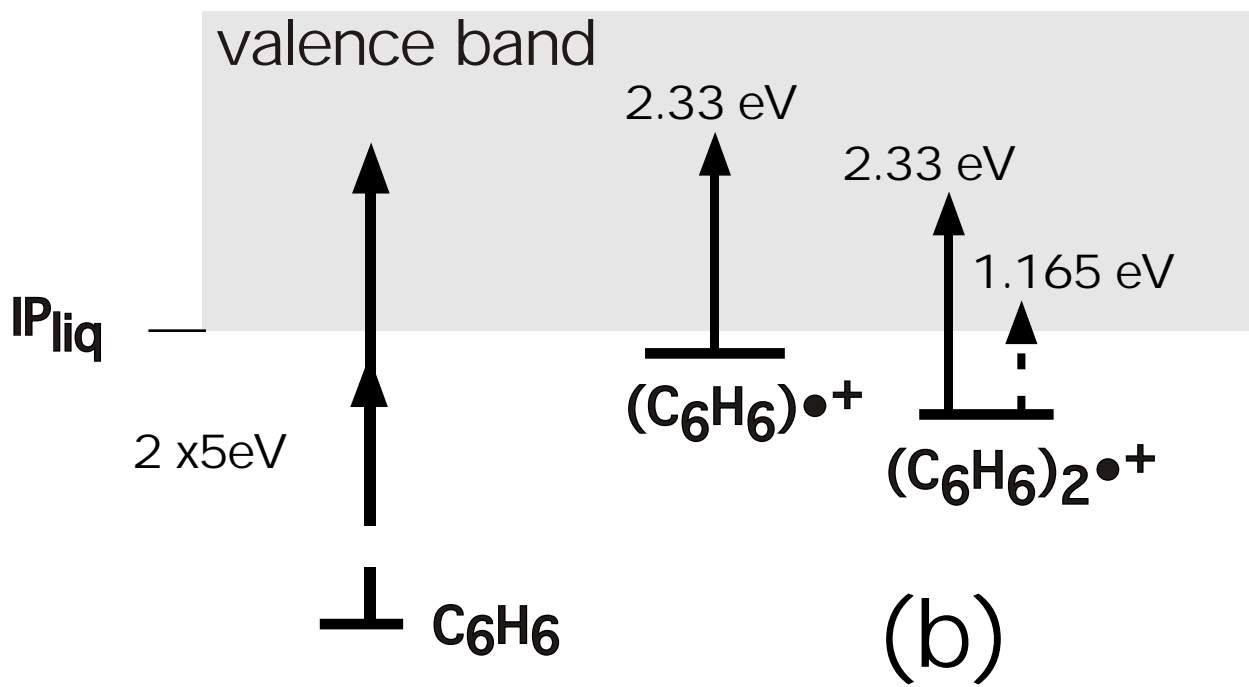

Figure 7